\documentclass[lettersize,journal]{IEEEtran}
\usepackage{amsmath,amsfonts}
\usepackage{algorithmic}
\usepackage{array}
\usepackage[caption=false,font=normalsize,labelfont=sf,textfont=sf]{subfig}
\usepackage{textcomp}
\usepackage{stfloats}
\usepackage{url}
\usepackage{verbatim}
\usepackage{graphicx}
\usepackage{cite}
\usepackage{balance}
\usepackage{xcolor}
\usepackage{tikz}
\usetikzlibrary{arrows.meta,positioning,calc}
\usepackage{listings}
\usepackage{placeins}
\usepackage{float}

\def\BibTeX{{\rm B\kern-.05em{\sc i\kern-.025em b}\kern-.08em
    T\kern-.1667em\lower.7ex\hbox{E}\kern-.125emX}}

\begin{document}

\title{CUDA Kernel Optimization and Counter-Free Performance Analysis for Depthwise Convolution in Cloud Environments}

\author{
Huriyeh Babak%
\thanks{Institute for Information Processing (TNT), Leibniz University Hannover, Germany. 
Email: huriyeh.babak@stud.uni-hannover.de}
\and
Melanie Schaller%
\thanks{Institute for Information Processing (TNT) and L3S Research Center, 
Leibniz University Hannover, Germany. 
Email: melanie.schaller@tnt.uni-hannover.de}
\thanks{Manuscript submitted to IEEE Transactions on Parallel and Distributed Systems (TPDS).}
}
\markboth{IEEE Transactions on Parallel and Distributed Systems}%
{Babak \MakeLowercase{\textit{et al.}}: CUDA Kernel Optimization and Counter-Free Performance Analysis for Depthwise Convolution in Cloud Environments}


\maketitle

\begin{abstract}
Efficient GPU execution of convolution operators is governed by memory-access efficiency, on-chip data reuse, and execution mapping rather than arithmetic throughput alone. This paper presents a controlled operator-level study of CUDA kernel optimization for the depthwise convolution used in Structured State Space Model Convolutional Diagonal (S4ConvD), together with a cloud-compatible, counter-free performance analysis methodology.

The operator, model, dataset, and training configuration are fixed, and only the CUDA kernel implementation is varied. The evaluated CUDA kernels comprise naive, global-memory-coalesced, shared-memory cache-blocked, and warp-tiled variants, covering forward, input-gradient, and weight-gradient execution paths under steady-state training conditions.

Performance is characterized using a counter-free methodology that combines CUDA-event timing, execution-path decomposition, analytically derived memory-traffic modeling, effective-bandwidth estimation, and roofline analysis. This enables profiling-like architectural insights without requiring hardware performance counters or privileged profiling access. The warp-tiled kernel reduces convolution runtime by $3.26\times$ relative to the naive CUDA baseline, while end-to-end training speedup reaches $1.29\times$. A PyTorch implementation is used separately for numerical validation and runtime context, but is not treated as a controlled architectural baseline.

Forward and input-gradient paths benefit substantially from improved locality and on-chip data reuse, whereas the reduction-dominated weight-gradient path remains the primary bottleneck. The results demonstrate that meaningful architecture-level GPU kernel analysis can be performed reproducibly in restricted cloud environments, even without access to hardware performance counters.
\end{abstract}
\section{Introduction}
\IEEEPARstart{P}{erformance} in GPU-accelerated systems is shaped not only by algorithmic complexity, but also by how computations are mapped to the GPU execution model and memory hierarchy. For memory-sensitive operators, runtime is often dominated by memory-access patterns, thread mapping, and on-chip data reuse rather than peak arithmetic throughput \cite{owens2008gpu, williams2009roofline, cudaProgrammingGuide, cudaBestPracticesGuide}.

Recent hardware-aware operators such as FlashAttention and Mamba show that substantial gains can be achieved by reducing data movement and improving locality \cite{dao2022flashattention, gu2024mamba}. However, the behavior of individual GPU kernels remains less well characterized under controlled conditions, especially across forward and backward execution. Backward paths often contain reduction-dominated computations that introduce synchronization and accumulation overhead, causing optimization effectiveness to differ across execution paths \cite{harris2007reduction, jia2018volta, markidis2018nvidia}.

Cloud-based GPU platforms such as Kaggle, Google Colab, and AWS further complicate performance analysis because access to low-level hardware counters and profiling tools such as Nsight Compute is often restricted. This raises a practical question: how much architecture-level performance insight can be recovered without hardware-level profiling support? As such platforms become increasingly common, reproducible analysis methods that remain effective under these constraints are needed.

In this work, we address this question through a controlled operator-level study of the depthwise convolution in Structured State Space Model Convolutional Diagonal (S4ConvD) \cite{schaller2025s4convd, gu2022dss}. The operator, model, dataset, and training configuration are fixed, while four CUDA kernel variants are evaluated: naive, global-memory-coalesced, shared-memory cache-blocked, and warp-tiled. Forward, input-gradient, and weight-gradient paths are analyzed separately to expose execution-path-specific bottlenecks.

\begin{figure}[t]
    \centering
    \includegraphics[width=\linewidth]{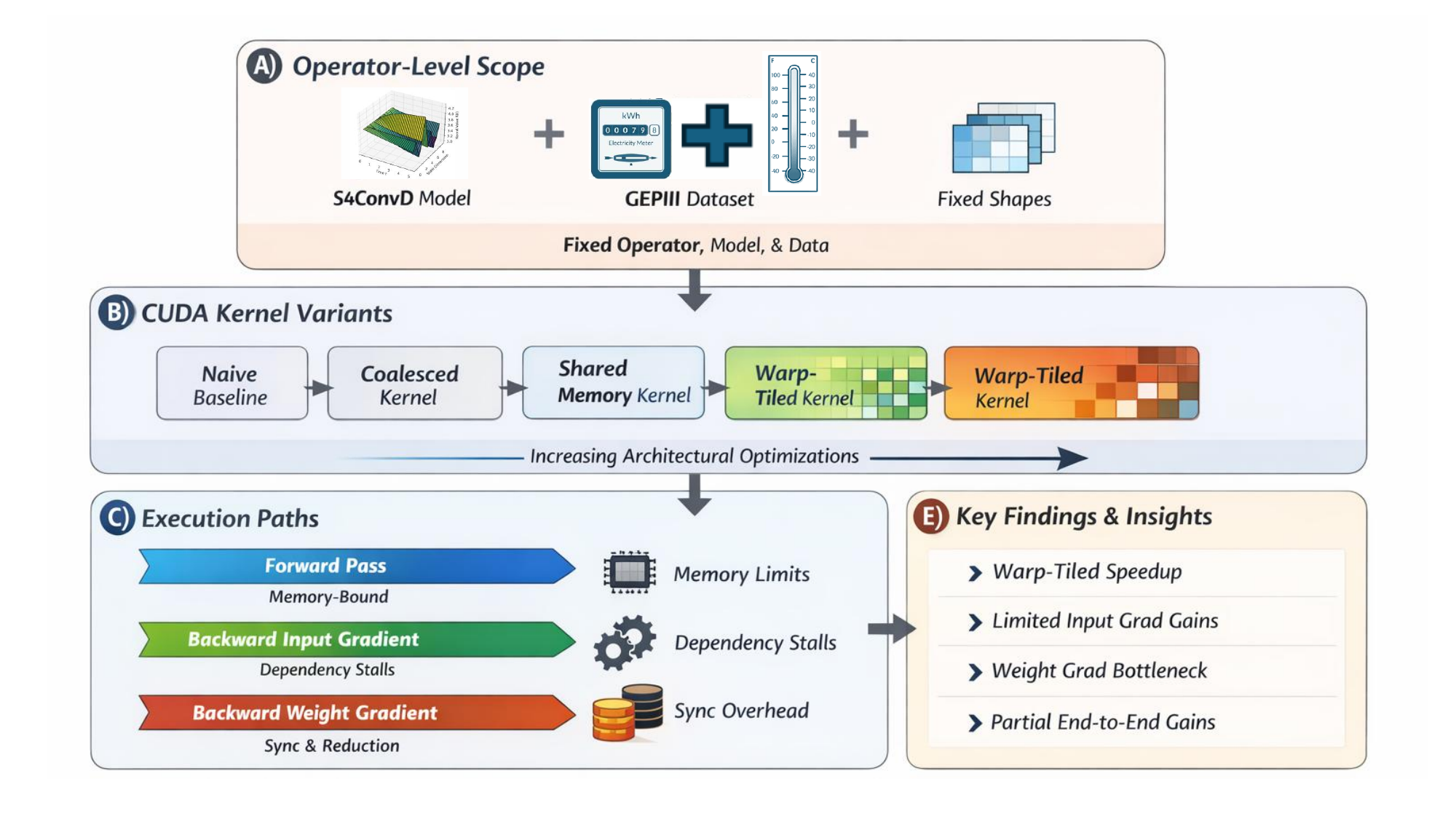}
    \caption{Conceptual overview of the study design and execution-path bottlenecks in depthwise convolution kernels. The figure illustrates the fixed operator-level scope, evaluated CUDA kernel variants, distinct execution paths, and the resulting architectural bottlenecks that govern performance.}
    \label{fig:overview}
\end{figure}

Fig.~\ref{fig:overview} illustrates the study design and the interaction between operator structure, kernel variants, and execution paths.

The main contributions of this work are as follows:

\begin{itemize}
    \item A cloud-compatible, counter-free methodology for GPU kernel analysis that reconstructs architecture-level performance characteristics using only portable runtime measurements and analytical modeling. The approach integrates CUDA-event timing, execution-path decomposition, memory-traffic estimation, effective-bandwidth analysis, and roofline modeling into a unified workflow, enabling profiling-like insights without hardware performance counters.

    \item A controlled operator-level evaluation framework that isolates the impact of CUDA kernel implementation by fixing the operator, model, dataset, and training configuration, enabling direct attribution of performance differences to execution mapping and memory-hierarchy utilization.

    \item An execution-path-aware characterization of depthwise convolution that systematically distinguishes throughput-oriented forward and input-gradient computations from the reduction-dominated weight-gradient path, revealing different optimization limits across execution paths.

    \item A quantitative analysis linking kernel design to memory traffic, effective bandwidth, and end-to-end performance, demonstrating that reducing redundant data movement yields substantially larger gains than access alignment alone, and explaining the non-linear translation from kernel-level acceleration to training speedup.
\end{itemize}

Although the individual CUDA optimization techniques are well established, the novelty of this work lies in the unified, execution-path-aware, and counter-free analysis methodology. The study links kernel design, analytical memory traffic, effective bandwidth, roofline behavior, and end-to-end training impact, showing that meaningful architectural insights can be obtained even in restricted cloud environments. The implementation and validation code are available online.\footnote{Code repository: \url{https://github.com/HuriyehBabak/CUDA_Kernels_S4ConvD}}
\section{Related Work}

This work relates to architecture-aware GPU kernel optimization, structured operators on parallel hardware, and performance limits caused by memory traffic and reductions. Unlike model-centric studies that emphasize end-to-end acceleration, we focus on controlled operator-level analysis of CUDA kernel behavior.

\subsection{Architecture-Aware GPU Kernel Optimization}

GPU performance depends on the interaction between SIMT execution, warp scheduling, and the hierarchical memory system \cite{owens2008gpu, kirk2016programming, cudaProgrammingGuide, cudaBestPracticesGuide, jia2018volta, markidis2018nvidia, volkov2008benchmarking}. Optimization strategies such as memory coalescing, shared-memory tiling, register blocking, and warp-centric execution improve bandwidth utilization, latency hiding, and data locality \cite{cudaBestPracticesGuide, harris2013optimizingcuda, boehm2022matmul}. 

Highly optimized libraries such as cuBLAS and cuDNN achieve near-peak performance for compute-intensive workloads \cite{chetlur2014cudnn, cublasMatmulGuide}. For memory-bound operators with limited data reuse, however, performance is primarily constrained by data movement and synchronization rather than arithmetic throughput \cite{williams2009roofline, mei2016memoryhierarchy}.

\subsection{Structured Operators and Reduction Constraints}

Structured sequential and convolutional operators benefit from hardware-aware design, as shown by FlashAttention, Mamba, and efficient convolutional architectures \cite{dao2022flashattention, gu2024mamba, gu2022dss, zhang2018shufflenet, howard2017mobilenets}. These approaches highlight the importance of IO-awareness and data reuse \cite{williams2009roofline}.

However, prior work largely focuses on fused operators or end-to-end acceleration rather than isolated kernel analysis. The Roofline model identifies memory bandwidth as the key limitation for low-arithmetic-intensity workloads \cite{williams2009roofline}, while backward execution introduces reduction-related synchronization and aggregation costs \cite{harris2007reduction, jia2018volta, markidis2018nvidia}. Reduction operations further introduce numerical sensitivity due to accumulation order \cite{higham2002accuracy}, leading to less favorable scaling compared to forward computation.

\subsection{Positioning of This Work}

In contrast to prior studies focusing on end-to-end acceleration or fused operator design \cite{chetlur2014cudnn, dao2022flashattention, gu2024mamba}, this work presents a controlled operator-level evaluation of CUDA kernels without modifying the underlying model. Using the depthwise convolution in S4ConvD \cite{gu2022dss, schaller2025s4convd}, we analyze forward, input-gradient, and weight-gradient execution.

By isolating kernel implementations, the study reveals operator-specific bottlenecks and shows how memory-access patterns, data reuse, and reduction structure govern performance beyond end-to-end runtime observations. 
Unlike prior work that relies on hardware performance counters or vendor-specific profiling tools, this work demonstrates that comparable architectural insights can be obtained using only portable runtime measurements and analytical modeling in restricted environments.
\section{Experimental Methodology and Evaluation Setup}
CUDA kernel variants are evaluated under controlled conditions using a cloud-compatible, counter-free methodology that combines CUDA-event timing, runtime decomposition, and analytical modeling to characterize kernel behavior without hardware performance counters \cite{cudaBestPracticesGuide, mei2016memoryhierarchy, williams2009roofline, jia2018volta}. The operator, model, dataset, and training configuration are fixed to ensure comparability with prior Structured State Space Model Convolutional Diagonal (S4ConvD) work \cite{schaller2025s4convd} and to isolate the impact of CUDA kernel implementation.

\subsection{Dataset and Workload}

Experiments use the ASHRAE Great Energy Predictor III (GEPIII) dataset \cite{miller2020ashrae}, which contains hourly energy consumption and meteorological features. Its low feature dimensionality and fixed sequence length make the depthwise convolution the dominant computational component, enabling controlled kernel-level analysis.

\subsubsection{Input Representation}

For each building $b_i$ and timestep $t_j$, the input vector is
\begin{equation}
\mathbf{u}^{(i)}_{t_j}
=
\bigl[
R^{(i)}_{t_j},
T^{(i)}_{a,t_j},
CC^{(i)}_{t_j},
T^{(i)}_{d,t_j}
\bigr]^{\top},
\label{eq:input_vector}
\end{equation}
where $R$ denotes energy consumption and $T_a$, $CC$, and $T_d$ denote meteorological variables \cite{miller2020ashrae}. All experiments use sequence length $L=48$ and input dimension $F=4$.

\subsection{Model Configuration}
\label{subsec:model-config}

The evaluated model is S4ConvD, based on diagonal state-space sequence modeling \cite{gu2022efficiently, gu2022dss, schaller2025s4convd}. Inputs are projected to latent dimension $H=128$ and processed by stacked S4ConvD blocks with nonlinear activation, channel-wise projection, and dropout rate 0.01. All architectural parameters are fixed across experiments.

\subsection{Training Configuration and Input Pipeline}

Training uses SGD with momentum 0.9, learning rate $10^{-3}$, gradient clipping with norm 1.0, RMSLE loss, and batch size $B=16{,}384$. Multi-worker loading and prefetching reduce data stalls so that measured runtime primarily reflects computation rather than input loading \cite{cudaBestPracticesGuide, paszke2019pytorch, lecun2015deep}.

\subsection{Hardware Platform}

Experiments are conducted on an NVIDIA Tesla P100-PCIE-16GB GPU based on the Pascal architecture \cite{nvidia2016p100datasheet, nvidia2016pascalwhitepaper}. Table~\ref{tab:hardware_platform} summarizes the hardware specifications relevant to CUDA parallelism, memory hierarchy, and resource constraints \cite{mei2016memoryhierarchy, cudaProgrammingGuide}.

\begin{table}[!t]
\caption{Experimental GPU Hardware Platform}
\label{tab:hardware_platform}
\centering
\footnotesize
\renewcommand{\arraystretch}{1.1}
\begin{tabular}{ll}
\hline
\textbf{Parameter} & \textbf{Specification} \\ \hline
GPU model & NVIDIA Tesla P100-PCIE-16GB \\
Architecture & Pascal \\
Compute capability & 6.0 \\
Streaming multiprocessors & 56 \\
Warp size & 32 threads \\
Max.\ threads per SM & 2048 \\
Max.\ threads per block & 1024 \\
Shared memory per SM & 64 KB \\
Max.\ shared memory per block & 48 KB \\
Registers per SM & 65{,}536 \\
L2 cache & 4 MB \\
Global memory & 16 GB HBM2 \\
Memory interface width & 4096 bit \\ \hline
\end{tabular}
\end{table}

\begin{figure}[!t]
\begin{center}
\includegraphics[width=\columnwidth]{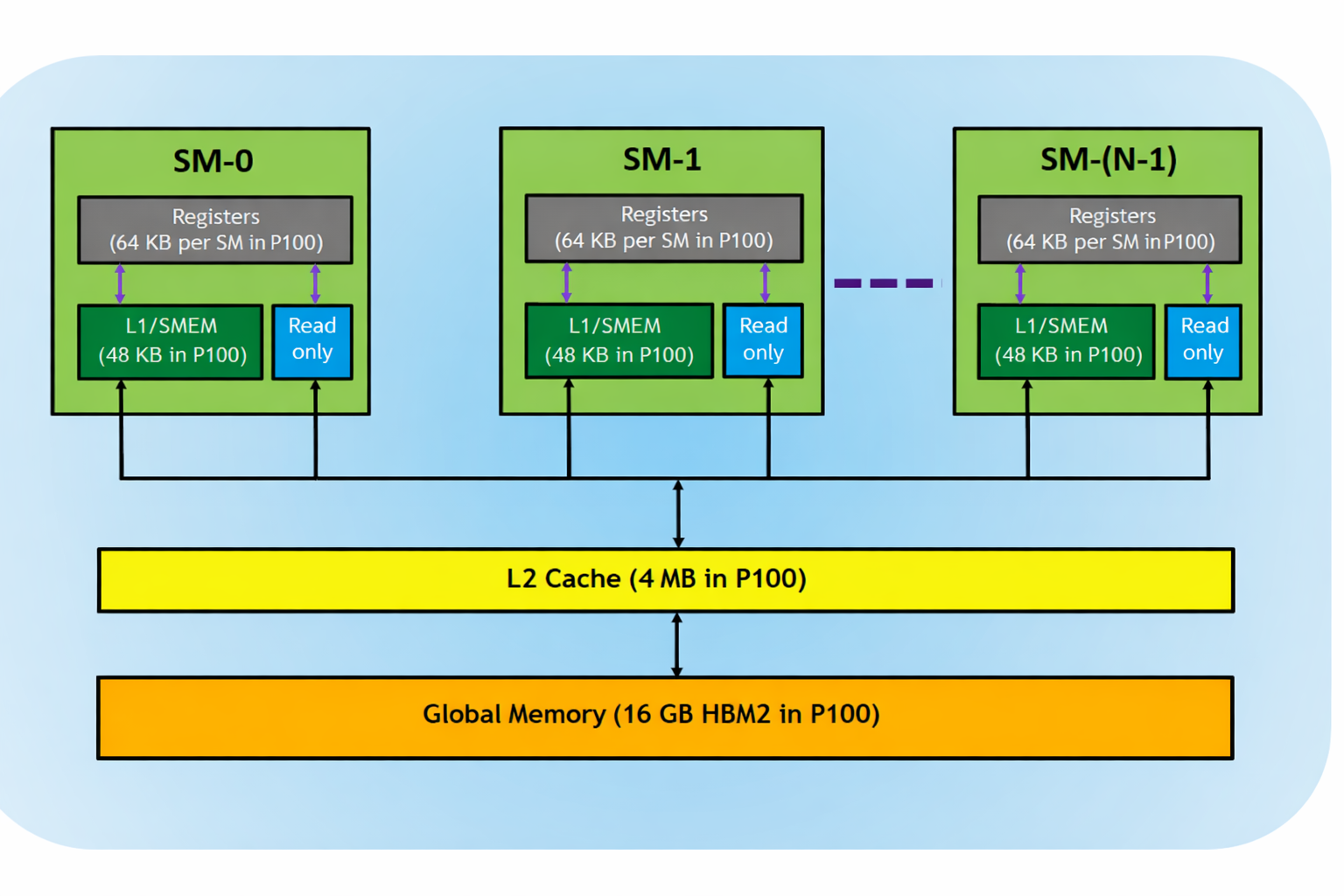}
\caption{Simplified memory hierarchy of the NVIDIA Tesla P100 used in the experiments \cite{nvidiaCudaRefresher}.}
\label{fig:p100_memory_hierarchy}
\end{center}
\end{figure}

Fig.~\ref{fig:p100_memory_hierarchy} illustrates the simplified memory hierarchy of the evaluated GPU.

\subsection{Numerical Validation}

All CUDA kernels are validated against a PyTorch reference implementation \cite{paszke2019pytorch}. Forward outputs and input gradients match within numerical precision. Weight gradients show small deviations due to floating-point accumulation order, as expected for parallel reductions, and do not affect training stability \cite{higham2002accuracy}.

\subsection{Performance Measurement}

Kernel runtime is the primary metric, complemented by epoch time and peak GPU memory usage. Runtimes are measured with CUDA events and explicit synchronization after warm-up \cite{cudaBestPracticesGuide}. Forward, input-gradient, and weight-gradient kernels are measured separately to expose execution-path-specific behavior.

We do not rely on PyTorch Profiler for execution-path decomposition. Instead, forward, input-gradient, and weight-gradient runtimes are measured explicitly using CUDA-event instrumentation, which keeps the methodology applicable in restricted cloud environments.

Measurements exclude data loading, optimizer updates, host-device transfers, and unrelated framework overhead where possible. Results are averaged over multiple runs, and steady-state training measurements exclude the warm-up epoch.

\subsection{Roofline Model Construction}
\label{subsubsec:roofline_method}

The roofline model is constructed from analytical operation counts, estimated data movement, and CUDA-event runtimes \cite{williams2009roofline, cudaBestPracticesGuide}. Arithmetic intensity is defined as floating-point operations per byte moved.

For forward and input-gradient computations, the operation count is
\begin{equation}
\text{FLOPs}_{\text{fwd/bwd\_in}}
=
B \cdot H \cdot L \cdot 2K,
\label{eq:flops_fwd_bwdin}
\end{equation}
where each multiply--add pair is counted as two floating-point operations. For the weight-gradient computation,
\begin{equation}
\text{FLOPs}_{\text{bwd\_k}}
=
H \cdot K \cdot B \cdot L \cdot 2.
\label{eq:flops_bwdk}
\end{equation}

Data movement is estimated from tensor sizes, access patterns, and kernel structure. Optimized kernels account for reduced redundancy from on-chip reuse, while the naive baseline uses logical data movement as a lower-bound proxy because redundant accesses depend on caching and scheduling behavior.

Achieved throughput and arithmetic intensity are computed as
\begin{equation}
\text{GFLOP/s}
=
\frac{\text{FLOPs}}{\text{runtime}},
\qquad
\text{AI}
=
\frac{\text{FLOPs}}{\text{bytes moved}}.
\label{eq:roofline_metrics}
\end{equation}

The memory and compute roofs use the P100 peak memory bandwidth of 732~GB/s and peak single-precision throughput of 10.6~TFLOP/s \cite{nvidia2016p100datasheet}. Small horizontal offsets are used only to avoid point overlap in the plot.

\subsection{Data Selection and Evaluation Protocol}

Development experiments use a reproducible 10\% subset with preserved temporal ordering. Since kernel runtime depends primarily on tensor dimensions, this subset is representative for implementation comparison. Final evaluation is performed on the full test set.

Forward and backward paths are analyzed separately to distinguish throughput-oriented from reduction-dominated workloads and to interpret the effects of coalescing, shared-memory reuse, and warp-level execution.
\section{Depthwise Convolution Operator and CUDA Kernel Variants}

This section presents CUDA kernel variants for the depthwise 1D convolution in Structured State Space Model Convolutional Diagonal (S4ConvD), designed to enable controlled, execution-path-aware characterization of kernel behavior. The variants range from a simple baseline to increasingly architecture-aware implementations based on global-memory coalescing, shared-memory cache blocking, and warp-tiled execution.

Across all variants, the mathematical operator remains unchanged; only execution mapping and memory-hierarchy utilization differ. This isolates the impact of thread-block configuration, memory-access organization, and on-chip data reuse on performance under hardware constraints such as register allocation and shared-memory usage \cite{cudaProgrammingGuide, cudaBestPracticesGuide}.

\begin{figure}[!t]
\centering
\includegraphics[width=\columnwidth]{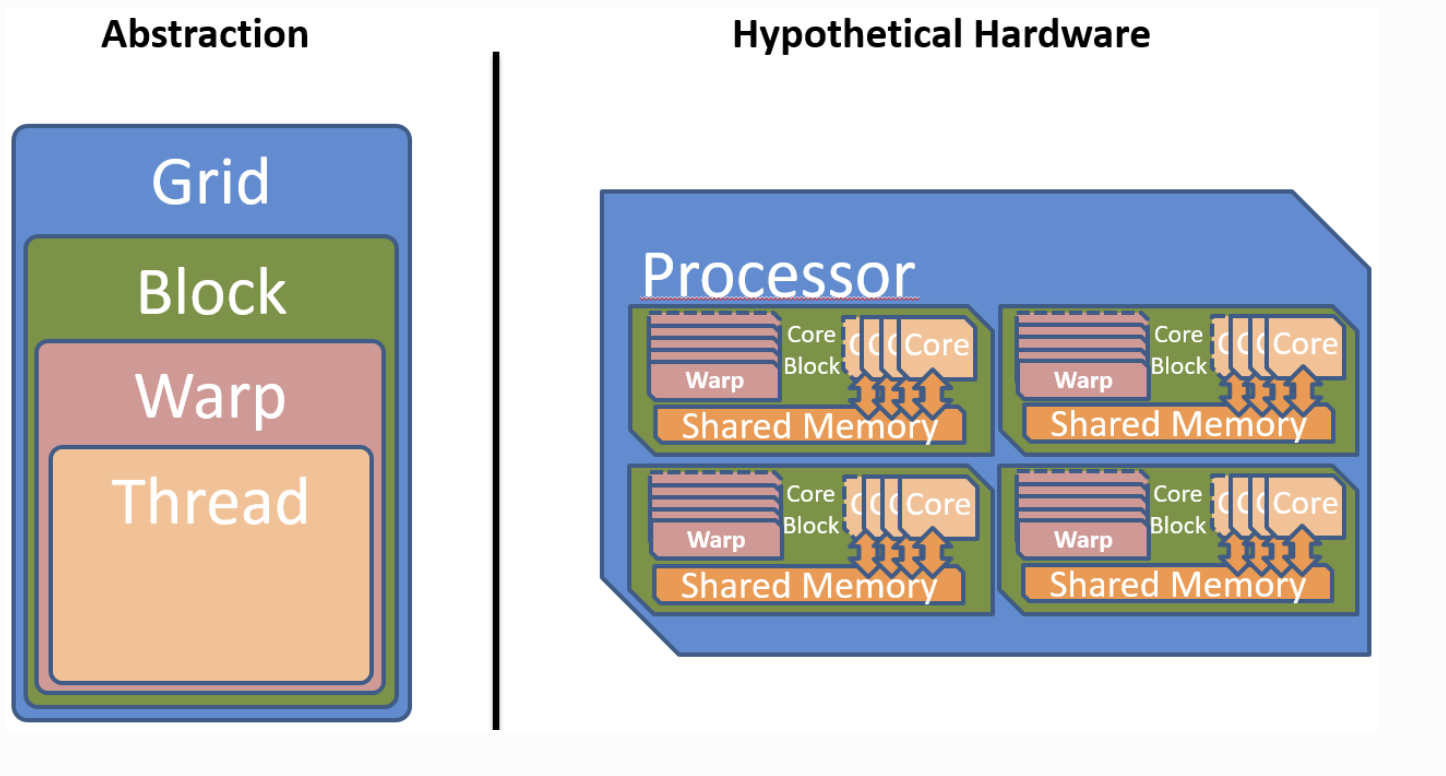}
\caption{Conceptual relationship between the CUDA execution hierarchy and the underlying GPU hardware organization. Thread blocks are scheduled onto streaming multiprocessors, where warps execute in parallel and share on-chip resources such as shared memory \cite{alpakaWarp}.}
\label{fig:cuda_abstraction}
\end{figure}

Performance is therefore governed not only by occupancy but also by memory-access efficiency, data reuse, synchronization overhead, and execution structure \cite{williams2009roofline, jia2018volta}. The resulting effects are analyzed quantitatively in Section~\ref{sec:results}. The relationship between the CUDA execution hierarchy and the underlying hardware organization is illustrated in Fig.~\ref{fig:cuda_abstraction}, which provides a conceptual reference for the execution mappings used in the following kernel designs.

\subsection{Common Problem Definition and Memory Layout}
\label{subsec:common-problem}

The depthwise convolution kernels operate on the projected latent representation
\begin{equation}
\mathbf{x}\in\mathbb{R}^{B\times H\times L},
\label{eq:latent_input}
\end{equation}
obtained after the input projection in Section~\ref{subsec:model-config}, where $B$, $H$, and $L$ denote batch size, number of channels, and sequence length, respectively. The corresponding kernel and output tensors are
\begin{equation}
\mathbf{k}\in\mathbb{R}^{H\times K}, \qquad
\mathbf{y}\in\mathbb{R}^{B\times H\times L},
\label{eq:kernel_output_shapes}
\end{equation}
with kernel length $K$. Since the convolution is depthwise, each channel $h$ is processed independently using a one-dimensional kernel $\mathbf{k}_{h,:}$, with no cross-channel interaction.

Let
\begin{equation}
p=\left\lfloor \frac{K}{2}\right\rfloor
\label{eq:padding_width}
\end{equation}
denote the padding width. The forward operator is defined as
\begin{equation}
y_{b,h,t}
=
\sum_{j=0}^{K-1}
\tilde{x}_{b,h,t+j-p}\,k_{h,j},
\label{eq:dwconv}
\end{equation}
where $\tilde{x}$ denotes zero-padded input. The corresponding input-gradient and kernel-gradient computations are
\begin{equation}
\nabla_{x_{b,h,t}}
=
\sum_{j=0}^{K-1}
\nabla_{y_{b,h,t+j-p}}\,k_{h,K-1-j},
\label{eq:dx_def}
\end{equation}
and
\begin{equation}
\nabla_{k_{h,j}}
=
\sum_{b=0}^{B-1}\sum_{t=0}^{L-1}
\nabla_{y_{b,h,t}}\,x_{b,h,t+j-p},
\label{eq:dk_def}
\end{equation}
respectively.

All tensors are stored in row-major order in global memory and processed in \texttt{float32} precision. For fixed $(b,h)$, the temporal index $t$ is the stride-1 dimension, so elements along $t$ are contiguous, and kernel weights are stored contiguously within each channel. This layout enables efficient stride-1 access and naturally supports coalesced global-memory transactions for adjacent threads \cite{cudaProgrammingGuide, cudaBestPracticesGuide}, forming the basis for the optimization strategies developed in the following subsections.
\subsection{Naive CUDA Baseline}

A simple CUDA baseline is implemented for the depthwise 1D convolution in \eqref{eq:dwconv}. Unlike optimized libraries such as cuDNN \cite{chetlur2014cudnn}, this implementation explicitly exposes thread mapping, memory-access behavior, and reduction structure, providing a transparent reference point for isolating the impact of subsequent execution-mapping and memory-hierarchy optimizations.

\begin{figure}[!t]
  \centering
  \includegraphics[width=\columnwidth]{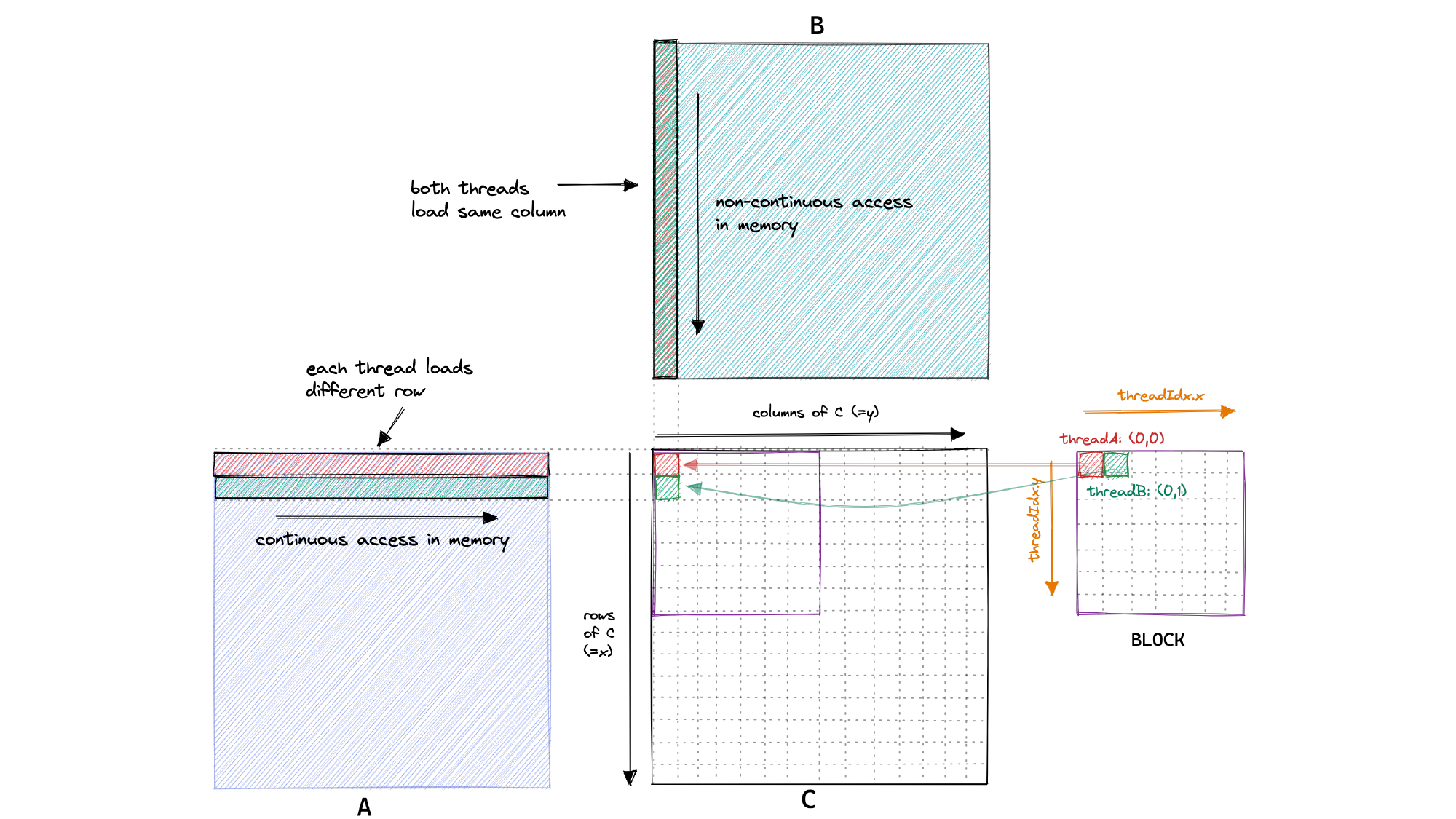}
  \caption{Naive CUDA parallelization strategy. Each thread computes one output element $y_{b,h,t}$, with the temporal dimension parallelized and the reduction over kernel width performed sequentially \cite{boehm2022matmul}.}
  \label{fig:naive-parallel}
\end{figure}

Fig.~\ref{fig:naive-parallel} illustrates the one-output-per-thread parallelization strategy. Each thread independently loads the input elements required for its convolution window and performs the reduction over the kernel width sequentially. Consequently, overlapping temporal regions lead to repeated global-memory accesses, and no parallelism is exposed across the reduction dimension.

\subsubsection{Forward Kernel}
\label{subsubsec:naive-forward}

The forward kernel assigns one output element to each thread according to
\begin{equation}
\begin{aligned}
b &= \texttt{blockIdx.z},\\
h &= \texttt{blockIdx.y},\\
t &= \texttt{blockIdx.x}\cdot\texttt{blockDim.x} + \texttt{threadIdx.x}.
\end{aligned}
\label{eq:naive_forward_map}
\end{equation}

One-dimensional thread blocks with 512 threads are used, resulting in a grid of size $\lceil L/512 \rceil \times H \times B$. This mapping exposes temporal parallelism without inter-thread cooperation, shared-memory staging, or on-chip data reuse.

\subsubsection{Backward Kernels}

The backward pass consists of input-gradient and kernel-gradient computations. The input-gradient kernel uses the same mapping as the forward pass and follows \eqref{eq:dx_def}, resulting in similar memory-access behavior.

The kernel-gradient computation follows \eqref{eq:dk_def}. Each thread is assigned to one coefficient $(h,j)$,
\begin{equation}
\begin{aligned}
h &= \texttt{blockIdx.y},\\
j &= \texttt{blockIdx.x}\cdot\texttt{blockDim.x} + \texttt{threadIdx.x},
\end{aligned}
\label{eq:naive_dk_map}
\end{equation}
and performs the accumulation over $B \cdot L$ sequentially. This design avoids atomic operations but does not expose parallelism across the reduction domain.

\subsubsection{Role as Reference Implementation}

The naive implementation serves as the primary CUDA baseline for all subsequent comparisons. Correctness is verified against a PyTorch reference implementation \cite{paszke2019pytorch}, while CUDA-event measurements provide the runtime reference for evaluating the optimized kernel variants.
\subsection{Optimization via Global-Memory Coalescing}

The naive baseline is dominated by redundant global-memory accesses and limited data reuse, making memory-access organization a primary optimization target \cite{cudaProgrammingGuide, cudaBestPracticesGuide}. This variant improves access efficiency at warp granularity without modifying the operator or introducing shared-memory staging.

The key idea is to align thread mapping with the memory layout such that consecutive threads access consecutive temporal elements. This enables coalesced global-memory transactions, reducing memory-transaction overhead and improving effective bandwidth utilization. While redundant loads across overlapping convolution windows remain, access efficiency is significantly improved.

\begin{figure}[!t]
\centering
\includegraphics[width=\columnwidth]{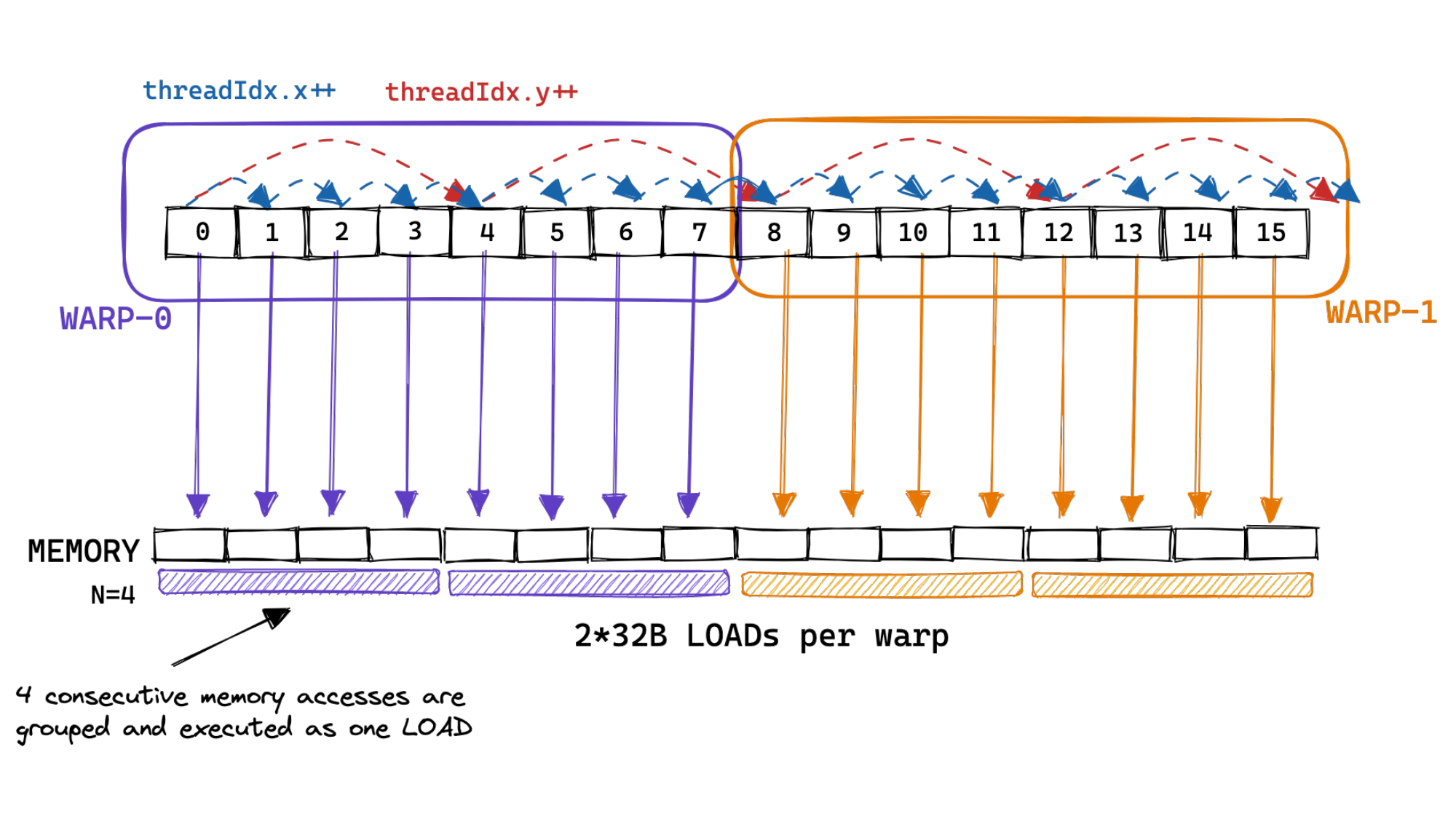}
\caption{Warp-level coalesced memory access. Consecutive threads access contiguous temporal elements, reducing the number of global-memory transactions \cite{boehm2022matmul}.}
\label{fig:coalescing}
\end{figure}

Fig.~\ref{fig:coalescing} illustrates the resulting warp-level access pattern.

\subsubsection{Forward}

The forward kernel uses one-dimensional thread blocks with 256 threads, organized as a tile over temporal and channel dimensions with $\texttt{TTILE}=32$ and $\texttt{HTILE}=8$. The mapping is
\begin{equation}
t = \texttt{blockIdx.x}\cdot\texttt{TTILE} + (\texttt{threadIdx.x}\bmod \texttt{TTILE}),
\end{equation}
\begin{equation}
h = \texttt{blockIdx.y}\cdot\texttt{HTILE} + \left\lfloor \frac{\texttt{threadIdx.x}}{\texttt{TTILE}} \right\rfloor,
\end{equation}
with $b = \texttt{blockIdx.z}$.

The launch configuration is
\begin{equation}
\texttt{grid} =
\left(
\left\lceil \frac{L}{\texttt{TTILE}} \right\rceil,\;
\left\lceil \frac{H}{\texttt{HTILE}} \right\rceil,\;
B
\right),
\quad
\texttt{block}=(256,1,1).
\end{equation}

Each thread computes one output element according to \eqref{eq:dwconv}. Because $\texttt{TTILE}=32$ matches the warp size, threads within a warp access contiguous memory locations, resulting in coalesced global-memory transactions. Kernel coefficients are accessed in a broadcast pattern and served efficiently by the cache hierarchy \cite{cudaProgrammingGuide}.

\subsubsection{Backward}

The input-gradient kernel uses the same mapping and exhibits similar access behavior.

The kernel-gradient computation remains reduction dominated. The reduction domain over $B\cdot L$ is partitioned into chunks processed independently by thread blocks. Each block computes partial sums using warp-level shuffle reduction, which are stored in an intermediate tensor and combined in a second reduction stage. This design avoids atomic operations while exposing parallelism across the reduction domain.

\subsubsection{Discussion}

Global-memory coalescing improves effective bandwidth utilization by aligning memory accesses with warp execution. However, since redundant data movement is not eliminated, the overall performance gain remains limited compared to later stages that introduce on-chip data reuse.
\subsection{Optimization via Shared-Memory Cache Blocking}

While global-memory coalescing improves access regularity, it does not eliminate redundant data movement, as overlapping temporal regions are still repeatedly loaded from global memory. This limits effective bandwidth utilization.

To address this limitation, this variant introduces shared-memory cache blocking. The key idea is to stage reusable input data and kernel coefficients on chip, enabling inter-thread data reuse and reducing redundant global-memory accesses.

\subsubsection{Forward}

The implementation operates on the tensors defined in Section~\ref{subsec:common-problem}. The batch and channel dimensions are flattened into a single index $s = bH + h$, preserving contiguous access along the temporal dimension.

Each thread block processes a temporal tile of length $TPB$, with mapping
\[
t = \texttt{blockIdx.x}\cdot TPB + \texttt{threadIdx.x}.
\]
An extended tile of size $TPB + K - 1$ is staged in shared memory to cover the convolution window, including halo elements.

Input data and kernel coefficients are cooperatively loaded into shared memory. After synchronization, each thread computes one output element using only shared-memory operands, eliminating redundant global-memory accesses within the convolution loop.

The shared-memory footprint is approximately 1404~B per block, well below hardware limits, allowing full occupancy while enabling efficient on-chip data reuse. Halo-loading overhead is amortized across threads.

\subsubsection{Backward}

The input-gradient computation follows the same tiling strategy and benefits from identical data reuse.

The kernel-gradient computation remains reduction dominated. The reduction domain is partitioned across thread blocks, which compute partial sums over subsets of the batch dimension. These partial results are stored and combined in a second reduction stage, avoiding atomic operations while exposing parallelism across the accumulation domain.

\subsubsection{Discussion}

Shared-memory cache blocking reduces redundant global-memory traffic by enabling on-chip reuse of overlapping temporal regions. This leads to a substantial increase in effective bandwidth utilization and performance for forward and input-gradient computations.

In contrast, the weight-gradient remains constrained by its reduction structure, highlighting the fundamental difference between throughput-oriented and reduction-dominated kernels.

Algorithmic changes to the reduction (e.g., hierarchical or fused approaches) are intentionally avoided to preserve comparability across kernel variants and isolate the impact of execution mapping and memory-hierarchy utilization.
\subsection{Warp-Tiled Execution}

This variant adopts a warp-centric design that maps one warp to a single $(b,h)$ instance. Because the temporal footprint fits entirely in shared memory, the complete working set can be staged on chip, enabling full data reuse without inter-warp coordination. This aligns the computation with the hardware warp abstraction and minimizes scheduling and synchronization overhead.

\subsubsection{Warp-Level Mapping}

A warp of $W=32$ threads is assigned to each $(b,h)$ pair, with
$b=\texttt{blockIdx.x}$, $h=\texttt{blockIdx.y}$, and
$\texttt{lane}=\texttt{threadIdx.x}$. Each lane computes up to two temporal positions,
\begin{equation}
t_0 = \texttt{lane}, \qquad
t_1 = \texttt{lane} + W,
\end{equation}
where $t_1$ is evaluated only if $t_1 < L$. This mapping covers the full temporal domain for one $(b,h)$ pair using a single warp.

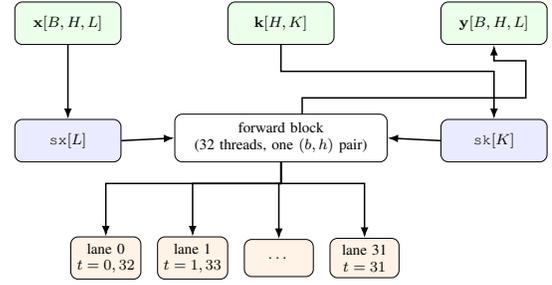
\begin{figure}[!t]
\centering
\resizebox{0.8\columnwidth}{!}{%
\begin{tikzpicture}[
    font=\footnotesize,
    box/.style={draw, rounded corners, minimum width=28mm, minimum height=9mm, align=center},
    gmem/.style={draw, rounded corners, fill=green!8, minimum height=8mm, minimum width=20mm, align=center},
    smem/.style={draw, rounded corners, fill=blue!8, minimum height=8mm, minimum width=20mm, align=center},
    lane/.style={draw, rounded corners, fill=orange!10, minimum height=7mm, minimum width=13mm, align=center},
    arrow/.style={-{Latex[length=2mm]}, thick}
]

\node[gmem] (x) {$\mathbf{x}[B,H,L]$};
\node[gmem, right=20mm of x] (k) {$\mathbf{k}[H,K]$};
\node[gmem, right=20mm of k] (y) {$\mathbf{y}[B,H,L]$};

\node[box, below=13mm of k, minimum width=40mm] (blk)
{forward block\\(32 threads, one $(b,h)$ pair)};

\node[smem, below=14mm of x] (sx) {$\texttt{sx}[L]$};
\node[smem, below=14mm of y] (sk) {$\texttt{sk}[K]$};

\node[lane, below=14mm of blk, xshift=-33mm] (l0) {lane 0\\$t=0,32$};
\node[lane, right=3mm of l0] (l1) {lane 1\\$t=1,33$};
\node[lane, right=3mm of l1] (ld) {$\cdots$};
\node[lane, right=3mm of ld] (l31) {lane 31\\$t=31$};

\draw[arrow] (x.south) -- ++(0,-5mm) -- (sx.north);
\draw[arrow] (k.south) -- ++(0,-5mm) -| (sk.north);

\draw[arrow] (sx.east) -- (blk.west);
\draw[arrow] (sk.west) -- (blk.east);

\draw[arrow] (blk.south) -- ++(0,-4mm) -| (l0.north);
\draw[arrow] (blk.south) -- ++(0,-4mm) -| (l1.north);
\draw[arrow] (blk.south) -- ++(0,-4mm) -| (ld.north);
\draw[arrow] (blk.south) -- ++(0,-4mm) -| (l31.north);

\draw[arrow]
  ($(blk.north)+(4mm,0)$)
  -- ++(0,3mm)
  -| ($(y.south)+(6mm,-3mm)$)
  -- ($(y.south)+(0,-3mm)$)
  -- (y.south);

\end{tikzpicture}%
}
\caption{Warp-centric mapping with full on-chip data reuse. Each warp processes one $(b,h)$ pair.}
\label{fig:warp_forward}
\end{figure}

This mapping eliminates inter-warp communication and reduces control divergence, enabling efficient warp-level execution.

\subsubsection{On-Chip Data Staging}

For each $(b,h)$ pair, the full input slice and kernel coefficients are staged in shared memory. All subsequent accesses are served from on-chip storage, eliminating repeated global-memory transactions.

The shared-memory footprint per block is
\begin{equation}
\texttt{SMEM}_{\mathrm{block}} = (L + K)\cdot 4~\text{bytes},
\end{equation}
which is approximately 384~B for the evaluated configuration. This small footprint enables high occupancy while maximizing data locality.

\subsubsection{Forward and Backward-Input Execution}

Both computations follow the same warp-level execution pattern. Each lane computes one or two output elements using shared-memory operands and fused multiply--add operations according to \eqref{eq:dwconv} and \eqref{eq:dx_def}. This design achieves full on-chip reuse and avoids redundant global-memory accesses.

\subsubsection{Backward Weight Gradient}

The kernel-gradient computation remains reduction dominated. The reduction is parallelized across thread blocks by partitioning the batch dimension. Each block stages inputs and gradients in shared memory, computes partial sums, and performs warp- and block-level reductions before writing results to global memory.

\begin{figure}[!t]
\centering
\resizebox{0.8\columnwidth}{!}{%
\begin{tikzpicture}[
    font=\footnotesize,
    box/.style={draw, rounded corners, minimum width=28mm, minimum height=9mm, align=center},
    smem/.style={draw, rounded corners, fill=blue!8, minimum width=16mm, minimum height=7mm, align=center},
    gmem/.style={draw, rounded corners, fill=green!8, minimum width=16mm, minimum height=7mm, align=center},
    redbox/.style={draw, rounded corners, fill=red!8, minimum width=28mm, minimum height=7mm, align=center},
    arrow/.style={-{Latex[length=2mm]}, thick}
]

\node[gmem] (gy) at (0,2.0) {$\nabla y$};
\node[gmem] (x)  at (3.6,2.0) {$x$};

\node[smem] (sgy) at (0,0.9) {$sGY$};
\node[smem] (sx)  at (3.6,0.9) {$sX$};

\node[box] (blk) at (1.8,-0.1) {block $(h,\mathrm{tile})$};
\node[redbox] (red) at (1.8,-1.3) {warp/block reduction};
\node[gmem] (gk) at (5.0,-1.3) {$\nabla k$};

\draw[arrow] (gy.south) -- (sgy.north);
\draw[arrow] (x.south) -- (sx.north);
\draw[arrow] (sgy.south) -- ++(0,-5mm) -- ($(blk.west)+(0,1.5mm)$);
\draw[arrow] (sx.south) -- ++(0,-5mm) -- ($(blk.east)+(0,1.5mm)$);
\draw[arrow] (blk.south) -- (red.north);
\draw[arrow] (red.east) --  (gk.west);

\end{tikzpicture}%
}
\caption{Warp-level reduction for kernel-gradient computation.}
\label{fig:warp_bwdk}
\end{figure}
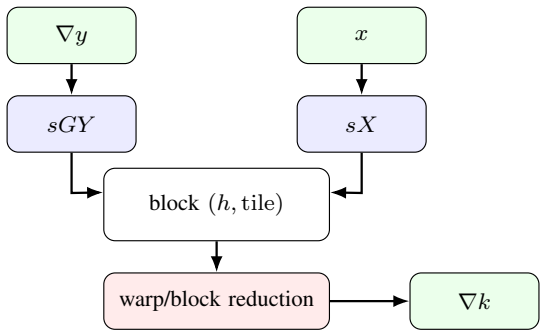

This design increases parallelism and reduces global-memory traffic, but synchronization and accumulation overhead remain due to the reduction structure.

\subsubsection{Discussion}

The warp-tiled design maximizes on-chip data reuse and aligns execution with warp granularity, reducing both memory traffic and scheduling overhead. As a result, forward and input-gradient computations achieve high efficiency. In contrast, the weight-gradient path remains constrained by its reduction-dominated structure.
\section{Experimental Results}
\label{sec:results}

This section evaluates the CUDA kernel variants and identifies the architectural factors governing their performance. The analysis combines CUDA-event timing, execution-path decomposition, analytically derived memory-traffic modeling, and roofline analysis to relate observed speedups to memory-access efficiency, on-chip data reuse, and reduction structure.

All measurements are performed without access to hardware performance counters. Instead, the evaluation relies on a counter-free methodology based on portable runtime measurements and analytical modeling, enabling architecture-level interpretation in restricted cloud environments.
\subsection{Numerical Validation and Stability}
\label{subsec:numerical_validation}

All CUDA kernel variants are validated against a PyTorch reference implementation \cite{paszke2019pytorch} across multiple problem sizes, including the full training configuration $(B,H,L)=(16384,128,48)$.

Forward outputs match the reference within machine precision, and input gradients exhibit maximum absolute error below $10^{-7}$. Small deviations are observed in the weight-gradient computation due to differences in floating-point accumulation order inherent to parallel reductions. For the largest configuration, the maximum absolute error is $4.96 \times 10^{-4}$, corresponding to a relative error on the order of $10^{-6}$.

These deviations are consistent with floating-point non-associativity in finite-precision arithmetic \cite{higham2002accuracy} and are expected for parallel reduction patterns on GPUs. Importantly, they do not affect training convergence or numerical stability in practice.

Additional validation results for the warp-tiled kernel are provided in Appendix~\ref{app:warp_validation}.

\subsection{Cross-Variant Runtime Summary}
\label{subsec:results-summary}

Table~\ref{tab:s4convd_main} summarizes steady-state performance across the CUDA implementations. The PyTorch implementation is not used as a baseline for the controlled kernel comparison, but is included separately in Appendix~\ref{app:warp_validation} for numerical validation and runtime context. For completeness, its execution time was measured, but it is excluded from the main table since it relies on backend library implementations that are not directly controlled in this study.

The naive CUDA baseline requires 133.47~ms for the convolution operator and 44.82~s per epoch.

Global-memory coalescing reduces convolution time to 106.65~ms, corresponding to a $1.25\times$ speedup over the naive CUDA baseline, indicating that improved access alignment alone provides only limited benefit. In contrast, shared-memory cache blocking reduces runtime to 66.57~ms by enabling on-chip data reuse. The warp-tiled implementation achieves the best CUDA-kernel performance, reaching 40.99~ms, corresponding to a $3.26\times$ speedup over the naive CUDA baseline and demonstrating the combined effect of warp-aligned execution and full data staging.

\begin{table}[!t]
\centering
\caption{Steady-state S4ConvD training performance (Epochs~2--5, excluding warm-up).}
\label{tab:s4convd_main}
\small
\setlength{\tabcolsep}{3pt}
\renewcommand{\arraystretch}{1.1}
\begin{tabular}{lccccc}
\hline
\textbf{Method} & \textbf{FWD} & \textbf{BWD\_in} & \textbf{BWD\_k} & \textbf{Conv Total} & \textbf{Epoch} \\
\hline
Naive CUDA & 29.97 & 30.25 & 73.26 & 133.47 & 44.82 \\
GMC        & 28.23 & 28.78 & 49.64 & 106.65 & 40.31 \\
Shared     & 16.36 & 16.03 & 34.17 & 66.57  & 36.91 \\
Warp-tiled & 10.46 & 10.61 & 19.91 & 40.99  & 34.74 \\
\hline
\end{tabular}

\vspace{1mm}
\begin{minipage}{\columnwidth}
\footnotesize
Kernel runtimes are reported in milliseconds; epoch time is reported in seconds.
FWD denotes forward execution, BWD\_in input-gradient execution,
BWD\_k weight-gradient execution, GMC the global-memory-coalesced kernel,
and Shared the shared-memory cache-blocked kernel.
\end{minipage}
\end{table}

\paragraph{Architectural interpretation.}
The results reveal a clear transition from access optimization to data-movement reduction. Coalescing improves transaction efficiency at the warp level but does not eliminate redundant global-memory traffic caused by overlapping convolution windows. In contrast, shared-memory cache blocking and warp-tiled execution reduce this redundancy by enabling explicit on-chip reuse, leading to substantially larger performance gains.

Across all variants, the weight-gradient path remains the dominant bottleneck. This limitation arises from the reduction-dominated structure of the computation, where accumulation across the batch and temporal dimensions limits parallelism and introduces synchronization overhead, thereby constraining achievable throughput despite improved memory locality.

\subsubsection{Kernel-Level and End-to-End Runtime Contribution}

Kernel-level acceleration does not translate linearly into end-to-end training speedup. While the warp-tiled implementation reduces total convolution runtime from 133.47~ms to 40.99~ms, corresponding to a $3.26\times$ improvement over the naive CUDA baseline, epoch time decreases only from 44.82~s to 34.74~s, corresponding to a $1.29\times$ improvement.

This discrepancy indicates that, as the convolution kernels become faster, non-kernel components such as framework overhead, synchronization, memory management, optimizer updates, and remaining model operations account for an increasingly large fraction of the total runtime. Therefore, kernel-level optimization must be interpreted together with application-level measurements.

\subsubsection{Runtime Distribution and Speedup}
\label{subsubsec:runtime_distribution}

\begin{figure}[t]
\centering
\includegraphics[width=\columnwidth]{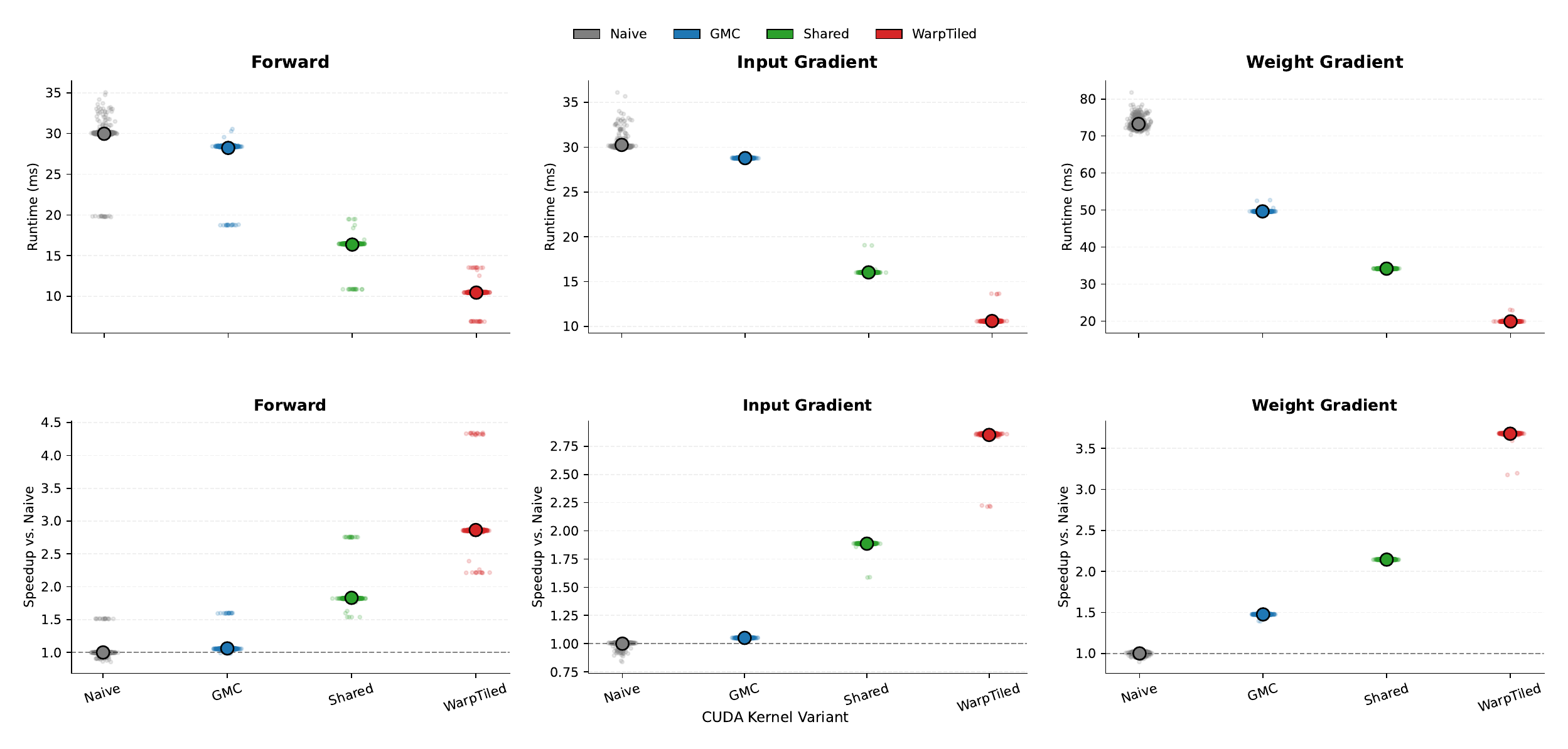}
\caption{Kernel runtime distribution and speedup across execution paths. 
The top row shows per-run runtime samples and mean runtimes for forward, 
input-gradient, and weight-gradient kernels. The bottom row shows the 
corresponding speedup relative to the naive CUDA baseline. Exact mean 
runtimes are reported in Table~\ref{tab:s4convd_main}.}
\label{fig:runtime_distribution}
\end{figure}

Table~\ref{tab:s4convd_main} reports absolute runtime values, while Fig.~\ref{fig:runtime_distribution} visualizes the corresponding per-path runtime reductions and speedups relative to the naive CUDA baseline. This separation keeps the numerical summary and visual analysis complementary.

Fig.~\ref{fig:runtime_distribution} shows that optimization effects are strongly dependent on the execution path. Forward and input-gradient computations benefit substantially from shared-memory cache blocking and warp-tiled execution, reaching approximately $2.9\times$ speedup over the naive CUDA baseline. The weight-gradient path achieves the largest relative speedup, $3.68\times$, but remains the slowest absolute component because it is dominated by reduction over the batch and temporal dimensions.

Global-memory coalescing provides only modest gains in forward and input-gradient execution because it improves access alignment without eliminating redundant loads across overlapping convolution windows. In contrast, shared-memory cache blocking and warp-tiled execution reduce redundant global-memory traffic by enabling explicit on-chip reuse. This explains the larger runtime reductions observed for these variants.

The figure also highlights why kernel-level acceleration does not translate directly into proportional end-to-end training speedup: even after optimization, the weight-gradient path remains a substantial component of convolution runtime, while non-kernel components increasingly account for the remaining epoch time.
\subsubsection{Counter-Free Effective Memory Bandwidth}
\label{subsubsec:effective_bandwidth}

Table~\ref{tab:effective_bw} reports counter-free effective memory-bandwidth estimates derived from CUDA-event runtimes and analytically modeled memory traffic. Since the experiments are conducted in a restricted cloud environment without access to hardware performance counters, these values should not be interpreted as direct DRAM-throughput measurements. Instead, they provide relative indicators of memory-access efficiency across the optimized kernel variants.

\begin{table}[t]
\centering
\caption{Counter-free effective memory-bandwidth estimates derived from CUDA-event timing and analytical memory-traffic modeling. Values indicate relative memory-access efficiency rather than direct hardware-counter measurements.}
\label{tab:effective_bw}
\footnotesize
\renewcommand{\arraystretch}{1.1}
\begin{tabular}{lcc}
\hline
\textbf{Variant} & \textbf{Eff. BW (GB/s)} & \textbf{Peak Util.} \\
\hline
Naive CUDA & N/A & N/A \\
GMC        & $\sim$42  & $\sim$6\% \\
Shared     & $\sim$75  & $\sim$10\% \\
Warp-tiled  & $\sim$115 & $\sim$16\% \\
\hline
\end{tabular}

\vspace{1mm}
\begin{minipage}{\columnwidth}
\footnotesize
The naive baseline is reported as N/A because its effective memory bandwidth cannot be estimated reliably from logical traffic alone. Overlapping convolution windows generate redundant global-memory accesses, and the actual number of memory transactions depends on cache behavior and scheduling effects that are not observable without hardware counters.
\end{minipage}
\end{table}

\begin{figure}[t]
\centering
\includegraphics[width=\columnwidth]{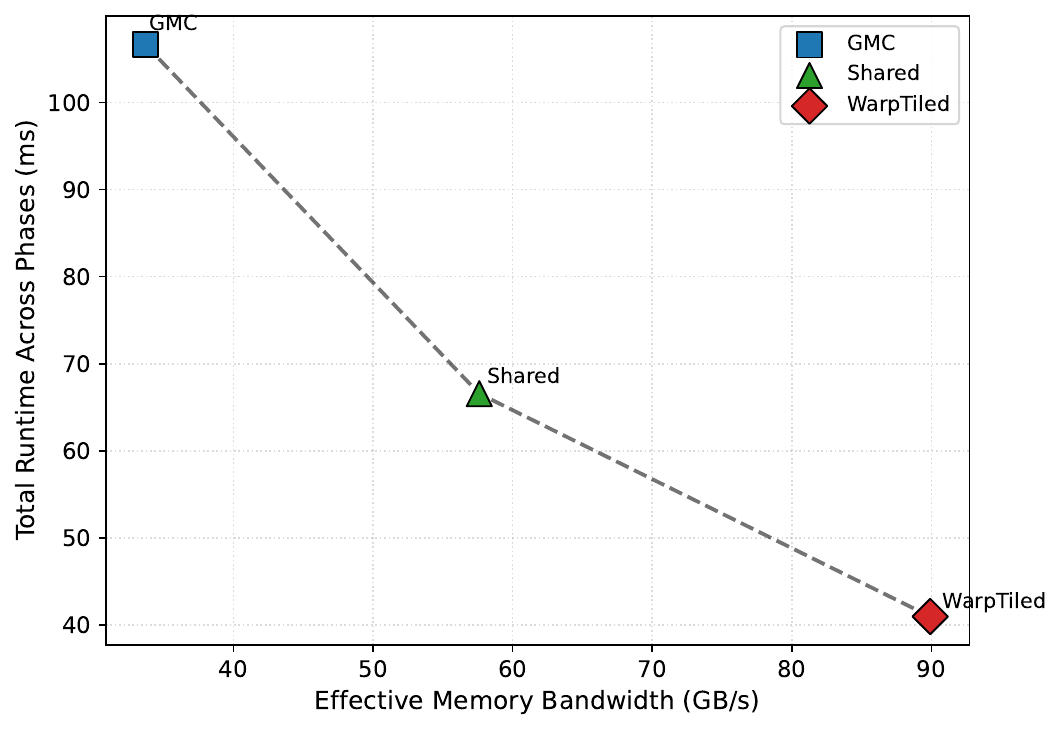}
\caption{Runtime bandwidth relationship for the optimized CUDA variants. The naive baseline is excluded because its redundant global-memory transactions cannot be reliably quantified without hardware counters. The inverse trend indicates that higher counter-free effective bandwidth corresponds to lower total convolution runtime.}
\label{fig:runtime_vs_bandwidth}
\end{figure}

The estimated effective bandwidth increases from the global-memory coalescing kernel to the warp-tiled implementation. This trend is consistent with the observed reduction in total convolution runtime and indicates that performance is governed primarily by effective data movement rather than peak arithmetic throughput.

Fig.~\ref{fig:runtime_vs_bandwidth} visualizes the relationship between effective bandwidth and total convolution runtime for the optimized CUDA variants. The naive baseline is excluded from this plot because its effective memory bandwidth cannot be estimated reliably without hardware counters. In the naive implementation, overlapping convolution windows generate substantial redundant global-memory accesses, while the realized number of memory transactions depends on cache behavior and scheduling effects that are not directly observable in the target environment.

Across the optimized variants, an inverse relationship between runtime and effective bandwidth is visible: higher effective bandwidth corresponds to lower total convolution runtime. Global-memory coalescing improves transaction efficiency but does not eliminate redundant data movement, resulting in only moderate gains. Shared-memory cache blocking and warp-tiled execution increase effective bandwidth more substantially by enabling on-chip data reuse and reducing global-memory traffic.

This interpretation is consistent with the runtime distributions in Fig.~\ref{fig:runtime_distribution}, where forward and input-gradient kernels benefit strongly from improved locality, while the weight-gradient path remains the dominant contributor due to its reduction-dominated structure. The roofline analysis in Fig.~\ref{fig:roofline} further confirms that all variants remain in the memory-bound regime, indicating that performance improvements arise from reduced data movement rather than increased computational throughput.

Although the estimated bandwidth values remain well below the theoretical peak of the P100 GPU, this gap is expected for low-arithmetic-intensity workloads with short sequence length, boundary handling, and reduction overhead. Therefore, the key result is not the absolute bandwidth value, but the consistent trend across kernel variants: reducing redundant data movement has a substantially larger impact than improving access alignment alone.

Overall, the results show that dominant architectural bottlenecks can be identified even without hardware performance counters. The combination of CUDA-event timing and analytical memory-traffic modeling is sufficient to reveal inefficient data movement as the primary performance limitation in this workload.
\subsubsection{Roofline Analysis}

\begin{figure}[!t]
\centering
\includegraphics[width=\columnwidth]{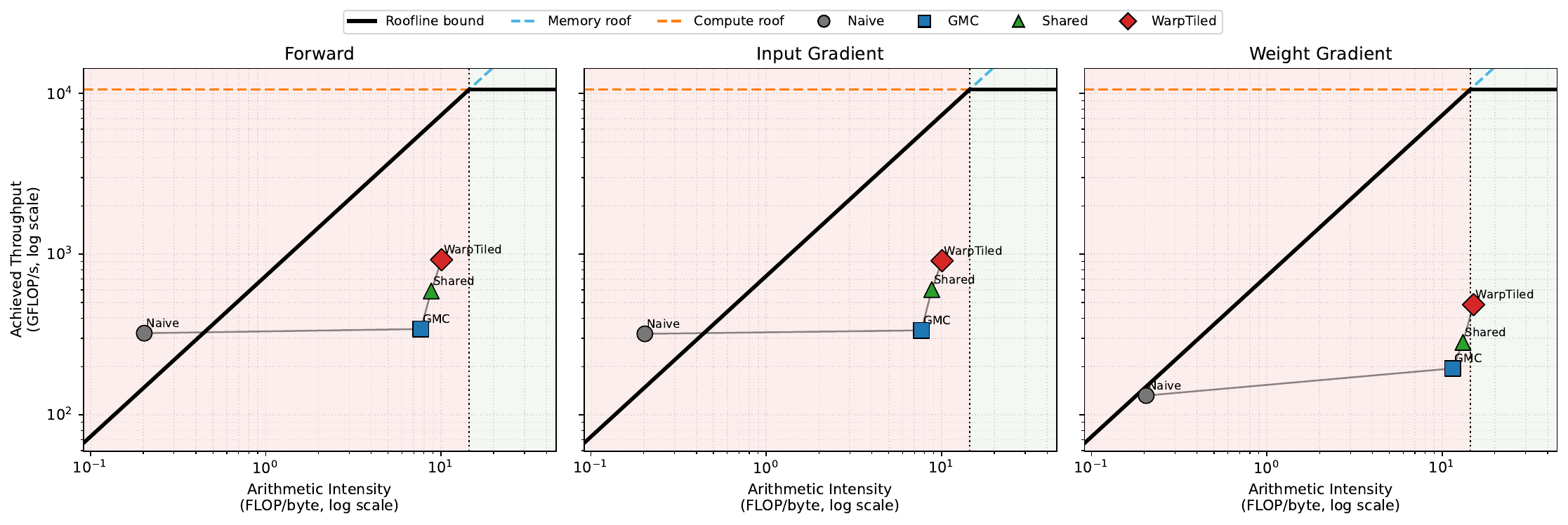}
\caption{Roofline analysis of CUDA kernel variants across forward, input-gradient, and weight-gradient execution paths.}
\label{fig:roofline}
\end{figure}

Fig.~\ref{fig:roofline} presents the counter-free roofline analysis of the evaluated CUDA kernel variants across all execution paths. The roofline model relates achieved throughput to arithmetic intensity and provides a compact visualization of whether performance is limited by memory bandwidth or compute throughput.

All kernel variants lie well below the compute roof and remain in the memory-bound region. The naive CUDA implementation exhibits low arithmetic intensity due to redundant global-memory accesses across overlapping convolution windows, which results in inefficient utilization of memory bandwidth.

Global-memory coalescing improves access regularity but does not significantly change arithmetic intensity, as redundant data movement remains largely unchanged. In contrast, shared-memory cache blocking and warp-tiled execution increase arithmetic intensity by enabling explicit on-chip data reuse. This shifts the kernels upward and slightly to the right in the roofline plot, reflecting improved bandwidth utilization.

Despite these improvements, none of the variants approach the compute roof, confirming that compute throughput is not the limiting factor and that the depthwise convolution remains memory-bound under the evaluated configuration. This observation is consistent with the effective-bandwidth analysis in Section~\ref{subsubsec:effective_bandwidth}, where performance improvements are primarily attributed to reduced redundant data movement.

The weight-gradient kernel achieves the largest relative speedup, but remains the slowest absolute component. This indicates that the optimizations reduce memory traffic, while the reduction-dominated structure continues to impose synchronization and accumulation overhead. As a result, even with improved data locality, its position in the roofline plot remains constrained compared to the forward and input-gradient paths.

Overall, the roofline analysis confirms that performance gains are driven by improved data reuse and reduced memory traffic rather than increased computational throughput. Importantly, this characterization is obtained without hardware performance counters, demonstrating that counter-free analysis is sufficient to capture the dominant performance behavior in restricted cloud environments.
\subsubsection{Implications for Memory-Bound Operators}
\label{subsubsec:implications}

The results have broader implications beyond the evaluated S4ConvD operator. Across all kernel variants, performance is governed primarily by data movement rather than arithmetic throughput, as consistently indicated by both the roofline analysis and effective-bandwidth trends.

This behavior is characteristic of memory-bound operators with low arithmetic intensity, where redundant memory traffic and limited on-chip reuse dominate execution cost. In such settings, improving memory-access alignment alone, for example through global-memory coalescing, provides only limited benefit. In contrast, reducing total data movement through shared-memory reuse yields substantially larger gains.

The observed progression from the naive baseline to the warp-tiled implementation shows that performance improvements are strongly correlated with the degree of data reuse and locality. Although demonstrated on S4ConvD, this trend is expected to generalize to other depthwise and channel-wise operators with similar access patterns, including depthwise convolutions in CNNs and structured state-space models.

At the same time, the persistent dominance of the weight-gradient kernel highlights a fundamental limitation of reduction-dominated workloads. Even with improved memory locality, large-scale accumulation introduces synchronization and serialization overhead that limits scalability on SIMT architectures.

These observations indicate that further performance improvements require algorithmic restructuring, such as more efficient reduction schemes or kernel fusion, rather than additional low-level memory-access optimizations alone.

This suggests that optimization efforts for similar operators should prioritize data reuse over purely access-alignment strategies.
\subsection{Counter-Free Performance Characterization}
\label{subsec:counter_free_characterization}

A central contribution of this work is to demonstrate that architecture-level performance insights can be obtained without access to hardware performance counters. The proposed methodology combines CUDA-event timing, execution-path decomposition, analytical memory-traffic modeling, effective-bandwidth estimation, and roofline analysis.

Together, these components expose the same dominant performance mechanisms typically identified using hardware-counter-based profiling, including redundant data movement, limited on-chip reuse, reduction overhead, and the gap between kernel-level and end-to-end performance.

This approach enables reproducible and portable performance analysis in cloud-based environments where access to low-level profiling tools is restricted, while preserving architectural interpretability. Importantly, it shows that meaningful architectural insights can be derived without privileged access to GPU internals.
\subsubsection{Bottleneck Characterization}

The analysis identifies two dominant architectural bottlenecks. First, the naive and coalesced kernels are limited by redundant global-memory traffic caused by repeated accesses to overlapping convolution windows. Second, the weight-gradient path remains reduction dominated, requiring aggregation across large batch and temporal domains.

Shared-memory cache blocking addresses the first bottleneck by enabling explicit data reuse and reducing redundant global-memory accesses. Warp-tiled execution further improves locality, load balance, and warp-level data reuse. In contrast, the second bottleneck is structural and persists across all kernel variants.

\subsubsection{System-Level Implications}

As kernel efficiency improves, system-level overheads increasingly dominate runtime. This limits the translation of kernel-level speedups into end-to-end training gains and highlights the importance of evaluating both kernel-level and application-level performance.

The results suggest that further acceleration requires structural changes to reduction-dominated computations rather than additional improvements in memory-access organization alone.

\subsection{Main Empirical Findings}

\begin{itemize}
\item Memory-access alignment alone provides limited benefit for memory-bound operators.
\item On-chip data reuse is the dominant factor driving performance improvement.
\item Optimization effectiveness depends strongly on the execution path.
\item Kernel-level speedup translates sublinearly to end-to-end performance.
\item Counter-free analysis is sufficient to identify the dominant architectural bottlenecks in the evaluated cloud-based setting.
\end{itemize}
\section{Conclusion}

This paper presented a controlled operator-level study of CUDA kernel optimization for the depthwise convolution in Structured State Space Model Convolutional Diagonal (S4ConvD). By fixing the operator, model, dataset, and training configuration, the analysis isolates the impact of execution mapping, memory-access organization, and on-chip data reuse, enabling direct attribution of performance differences to architectural factors.

The results consistently show that performance is governed primarily by data movement rather than arithmetic throughput. Improving access regularity through global-memory coalescing yields only moderate gains, as it reduces transaction overhead but does not eliminate redundant data movement. In contrast, shared-memory staging and warp-aligned execution reduce the number of global-memory transactions per output element by enabling explicit data reuse, resulting in a $3.26\times$ kernel-level speedup over the naive CUDA baseline. The PyTorch reference implementation, reported separately for validation and runtime context, exhibits substantially higher convolution runtime but is not used as a controlled baseline due to its reliance on backend library optimizations.

A key observation is the strong asymmetry across execution paths. While forward and input-gradient computations benefit directly from improved locality and warp-level execution, the weight-gradient path remains constrained by its reduction-dominated structure. This structural bottleneck introduces synchronization and accumulation overhead that persists across all optimization stages and ultimately dominates overall runtime.

The study further shows that occupancy alone is not a reliable predictor of performance for memory-bound kernels. High-occupancy kernels can remain inefficient when memory accesses are redundant, whereas lower-occupancy kernels can achieve higher performance by reducing data movement and improving locality. This highlights the importance of memory efficiency over raw parallelism for memory-bound workloads.

Another key finding is the systematic gap between kernel-level acceleration and end-to-end performance. Even substantial reductions in convolution runtime translate only partially into training acceleration, as system-level overheads such as framework execution, memory management, and synchronization become increasingly dominant.

\paragraph{Implications.}
These results generalize to memory-bound operators with low arithmetic intensity and reduction-heavy execution patterns. In such workloads, reducing redundant data movement and increasing on-chip reuse are more effective than improving access alignment or increasing parallelism alone. The observed behavior is therefore expected to extend to other depthwise and channel-wise operators with similar memory-access characteristics.

\paragraph{Counter-free performance analysis.}
A central contribution of this work is to demonstrate that architecture-level performance bottlenecks can be identified without relying on hardware performance counters. By combining CUDA-event timing, execution-path decomposition, analytical memory-traffic modeling, effective-bandwidth estimation, and roofline analysis, the proposed methodology provides insights consistent with profiling-based analysis in restricted environments. The consistency between runtime trends, effective-bandwidth estimates, and roofline positioning further supports the validity of this counter-free approach.

Although the individual optimization techniques are well established, the main contribution lies in the unified, execution-path-aware analysis enabled by the proposed workflow, which links kernel design, data movement, and performance behavior under restricted conditions.

This partially reframes restricted cloud environments from a limitation into a reproducibility opportunity by enabling standardized, portable, and hardware-agnostic performance analysis workflows. Cloud-based GPU environments can reduce variability in the hardware class, driver stack, CUDA version, and software configuration, enabling more consistent experimental conditions than many heterogeneous local setups.

\paragraph{Future directions.}
The persistent dominance of the weight-gradient computation suggests that further performance improvements require algorithmic restructuring, such as more efficient reduction strategies or kernel fusion. Future work may extend the proposed counter-free methodology to more complex operators, multi-kernel pipelines, and automated analysis workflows, enabling broader adoption of reproducible GPU performance characterization in cloud environments.
\appendices
\section{Additional Numerical Validation of the Warp-Tiled Kernel}
\label{app:warp_validation}

As discussed in Section~\ref{subsec:results-summary}, the warp-tiled kernel achieves the best performance among the evaluated CUDA implementations. This appendix provides additional numerical validation and reference runtime context to confirm correctness and clarify the role of the PyTorch implementation.

\subsection{Validation Protocol}

The warp-tiled kernel is validated against a PyTorch grouped \texttt{conv1d} implementation, which serves as a reference for numerical correctness. It is important to note that this reference is used exclusively for validation purposes and is not part of the controlled kernel-level performance comparison.

Validation covers forward execution, input-gradient, and weight-gradient computations. Experiments span a range of configurations by varying batch size $B$, channel dimension $H$, sequence length $L$, and kernel size $K$. Smaller validation cases use different kernel sizes to test correctness across multiple shapes, while the main benchmark configuration uses $(B,H,L)=(16384,128,48)$ with a convolution kernel length of $K=48$.

For each configuration, outputs and gradients produced by the warp-tiled kernel are compared element-wise against the reference. The maximum absolute difference is reported for all quantities. For the weight-gradient path, which is most sensitive to accumulation order, the relative error is additionally evaluated.

For even kernel sizes such as $K=48$, the PyTorch reference uses zero padding of $K/2$, and the output is cropped to the input sequence length to match the custom CUDA kernel convention.

\subsection{Reference Runtime Context}

Although the PyTorch grouped \texttt{conv1d} implementation is used only as a numerical reference and is not included in the controlled CUDA-kernel comparison, its execution time was measured to provide additional runtime context.

For the full benchmark configuration $(B,H,L)=(16384,128,48)$ with $K=48$, the PyTorch reference requires 28.44~ms for the forward pass, 25.62~ms for the input-gradient computation, and 141.73~ms for the weight-gradient computation, resulting in a total convolution runtime of 195.79~ms.

These measurements are not used to attribute architectural effects, since the PyTorch implementation relies on backend library behavior that is not directly controlled in this study. Instead, they serve as a sanity-check reference and provide context for the scale of execution-path runtimes.

\subsection{Observed Numerical Behavior}

Across all tested configurations, forward outputs match the reference within float32 numerical precision, and input gradients remain numerically stable with negligible deviations.

The weight-gradient exhibits small differences that increase with problem size. This behavior is expected and results from variations in floating-point accumulation order inherent to parallel reduction on GPU architectures. For the largest configuration $(B,H,L)=(16384,128,48)$ with $K=48$, the maximum absolute difference is $4.96\times10^{-4}$, corresponding to a relative error of approximately $1.05\times10^{-6}$.

These deviations remain well within the tolerance of single-precision training workloads and are consistent with established numerical properties of floating-point reductions.

Overall, the results confirm that the warp-tiled implementation preserves numerical correctness across all execution paths.

\subsection{Error Trend Across Problem Sizes}

\begin{figure}[!t]
    \centering
    \includegraphics[width=\columnwidth]{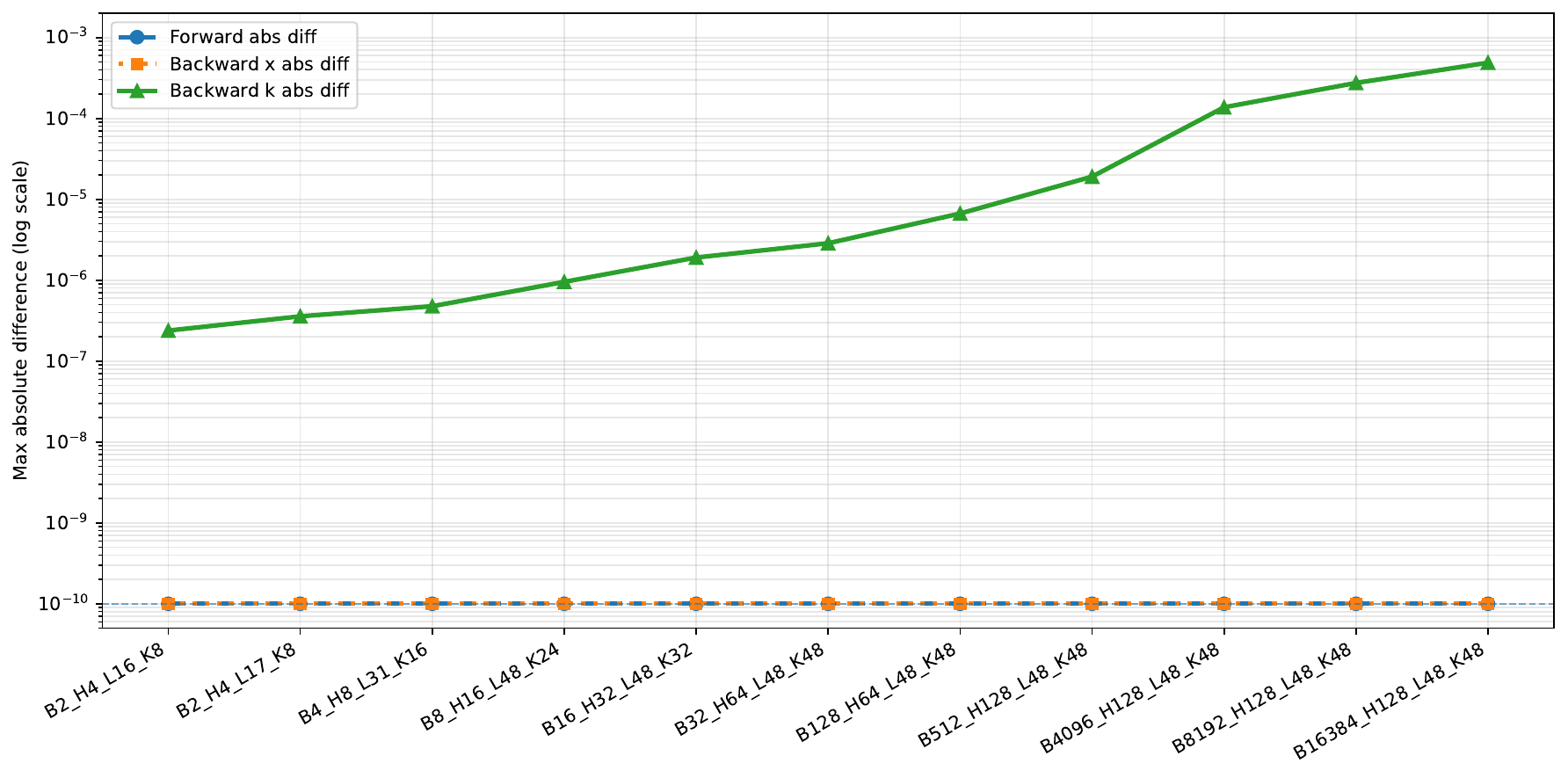}
    \caption{Maximum absolute difference between the warp-tiled implementation and the PyTorch reference across problem sizes. Forward and input-gradient errors remain at the numerical precision floor, while weight-gradient error increases gradually due to accumulation-order differences. Values below the plotting threshold are clipped only for visualization.}
    \label{fig:appendix_warp_validation}
\end{figure}

Fig.~\ref{fig:appendix_warp_validation} shows the maximum absolute differences for forward, input-gradient, and weight-gradient computations as a function of problem size.

Forward and input-gradient errors remain at the numerical precision floor across all configurations. In contrast, the weight-gradient error increases gradually with accumulation depth, reflecting the expected sensitivity of reduction-dominated computations to floating-point ordering effects.

For visualization on a logarithmic scale, values below a fixed threshold are clipped for display purposes only; all reported values are computed from the original outputs without modification.

\FloatBarrier

\subsection{Module-Level Validation}

In addition to operator-level validation, the warp-tiled kernel is evaluated within the full S4ConvKernel module forward path. For the tested configuration, the custom implementation produces outputs that match the reference implementation within float32 numerical precision, with no observable deviation in maximum absolute error.

This confirms that the kernel is not only correct in isolation but also integrates consistently within the model-level execution.

\subsection{Summary}

The presented validation results demonstrate that the warp-tiled kernel maintains numerical correctness across all execution paths and problem scales. The observed deviations in the weight-gradient are consistent with expected floating-point reduction effects and remain well within acceptable tolerance.

These findings confirm that the performance improvements achieved by the warp-tiled design do not compromise numerical stability, supporting the validity of the proposed counter-free performance analysis.

\bibliographystyle{IEEEtran}
\bibliography{references}

\begin{IEEEbiographynophoto}{Huriyeh Babak}
Huriyeh Babak is currently pursuing the M.Sc. degree in computer science at Leibniz-University Hannover, Hannover, Germany. Her research interests include high-performance computing, GPU programming, and deep learning systems.
\end{IEEEbiographynophoto}

\begin{IEEEbiographynophoto}{Melanie Schaller}
Melanie Schaller received the Ph.D. from the University of Würzburg in 2023 and worked as Researcher for the Center of Artificial Intelligence and Data Science (CAIDAS) at the University of Würzburg from 2021 till 2024. Her research interests include machine learning systems, anomaly detection in multivariate time series, graph signal processing, and sensor network applications as well as sequential modelling with deep state-space models. She has contributed to several research projects in machine learning for engineering purposes, including anomaly detection in structural health monitoring and leakage detection in water distribution networks. Since 2024 she works as a research group leader at the Institute for Information Processing (tnt) at Leibniz-University Hannover and is also a member of the Management Board of the L3S Research Center.
\end{IEEEbiographynophoto}

\end{document}